# Conductive Buffer Layers and Overlayers for the Thermal Stability of Coated Conductors

C. Cantoni, T. Aytug, D. T. Verebelyi, M. Paranthaman, E. D. Specht, D. P. Norton, and D. K. Christen

*Abstract*—We analyze fundamental issues related to the thermal and electrical stability of a coated conductor during its operation. We address the role of conductive buffer layers in the stability of Ni-based coated conductors, and the effect of a metallic cap layer on the electrical properties of Ni alloy-based superconducting tapes. For the first case we report on the fabrication of a fully conductive RABiTS architecture formed of bilayers of conductive oxides $SrRuO_3$ and $LaNiO_3$ on textured Ni tapes. For the second case we discuss measurements of current-voltage relations on $Ag/YBa_2Cu_3O_{7-\delta}$ and $Cu/Ag/YBa_2Cu_3O_{7-\delta}$ prototype multilayers on insulating substrates. Limitations on the overall tape structure and properties that are posed by the stability requirement are presented.

## I. INTRODUCTION

At the stage of implementation, coated conductor tapes will have to be electrically and thermally stabilized against transient conditions in which the current may exceed $I_c$ of the HTS coating. Coated conductors are typically formed of a 50 μm thick metallic substrate (Ni or Ni-alloys), one or more insulating oxide buffer layers with a total thickness of 0.5-1 μm, and an HTS film thicker than 1μm. As a benchmark we can assume that the HTS film (usually YBCO) carries a current density of $1\times10^6$ A/cm$^2$ in the superconducting state at liquid nitrogen temperatures (64 - 77 K) and in self magnetic field. In the event of a transient to the dissipative regime, the power generated per unit area in such an electrically-isolated 2 μm thick HTS layer would exceed $10^4$ W/cm$^2$, resulting in destruction of the superconductor.

In order to remove the heat generated from the HTS layer, and restore superconductivity, the superconducting film has to be electrically coupled to a conductive metallic layer. This coupling provides a lower overall resistivity during transient losses of superconductivity, and, consequently, a lower dissipation in the net conductor. In addition, coupling to a good metal greatly increases the thermal conductivity of the entire tape, significantly improving the outflow of heat to the surrounding refrigerant bath, and, therefore, the cooling rate of the conductor. The efficiency of cooling depends on many parameters of the application design. However, we consider the simple case of steady-state heat flow removed by contact of the tape with boiling liquid nitrogen. Then, the maximum temperature excursion is determined by limiting the heat flux to a value below the critical heat flux of $LN_2$ at 77 K (10-20 W/cm$^2$) [1].

Electrical coupling of the superconducting film to the metal layer can be envisioned in two different ways. The textured metal substrate of sufficiently low resistivity, as in the case of pure Ni or Cu, can itself provide the necessary stability for the HTS film if the buffer layer is conducting rather than insulating. For a 2 μm thick YBCO coating that is electrically connected to an underlying 50 μm thick base metal tape of pure nickel, the normal-state heat flux from each unit surface area would be only about 4 W/cm$^2$ per tape surface, well below the critical heat flux of liquid nitrogen.

On the other hand, a conductive buffer layer is not an effective route if the substrate of choice is a Ni alloy, which typically would have a resistivity comparable to the normal state resistivity of the superconductor (~100 μΩ-cm @ 100 K). In this case, stability can be achieved by applying a metal overlayer such as Cu or Ag on the superconducting film.

Figure 1 shows the calculated dependence on the substrate resistivity of the metal-layer thickness necessary to maintain the transient heat flux below 5 W/cm$^2$ when an HTS film 2 μm thick, carrying a current density of 1 MA/cm$^2$, is driven to the dissipative state. The effect on the overall tape current

C. Cantoni is with Solid State Division, Oak Ridge National Laboratory, Oak Ridge, TN 37831-6061 USA (telephone: 865-574-6264, e-mail: cantonic@ornl.gov).

T. Aytug is with Solid State Division, Oak Ridge National Laboratory, Oak Ridge, TN 37831-6061 USA (telephone: 865-574-6271, e-mail: aytug@solid.ssd.ornl.gov).

D. T. Verebelyi was with Solid State Division, Oak Ridge National Laboratory, Oak Ridge, TN 37831-6061. He is now with American Superconductor Corporation (telephone: 508-621-4368, e-mail: dverebelyi@amsuper.com).

M. Paranthaman is with Chemical and Analytical Sciences Division, Oak Ridge National Laboratory, Oak Ridge, TN 37831-6100 (telephone: 865-574-5045, e-mail: paranthamanm@ornl.gov).

E. D. Specht is with Metals and Ceramics Division, Oak Ridge National Laboratory, Oak Ridge, TN 37831-6118 (telephone: 865-574-7682, e-mail: spechted@ornl.gov).

D. P. Norton was with Solid State Division, Oak Ridge National Laboratory, Oak Ridge, TN 37831-6059. He is now with Department of Materials Science and Engineering, University of Florida, Gainesville, FL 32611-6400 (telephone: 352-846-0525, e-mail: dnort@mse.ufl.edu).

D. K. Christen is with Solid State Division, Oak Ridge National Laboratory, Oak Ridge, TN 37831-6061 USA (telephone: 865-574-6269, e-mail: christendk@ornl.gov).

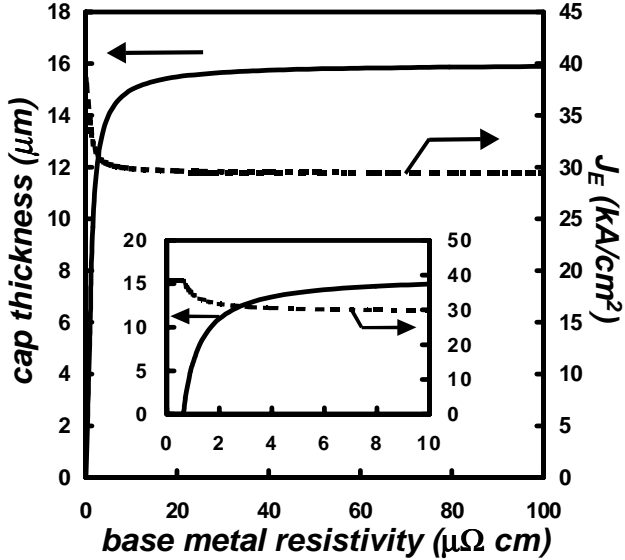

Fig.1. The approximate calculated thickness required of a highly conductive cap layer (Cu or Ag), in order to stabilize the transient dissipative heat flux to a value ~5 W/cm$^2$ (2.5 W/cm$^2$ per tape surface), as a function of the 50 μm thick substrate-tape resistivity. The HTS layer is taken to be 2 μm thick, carrying a current density of 1 MA/cm$^2$. The contact resistance between the HTS and the metal layer is assumed to be negligible. The inset shows the same plot expanded in the region of low resistivities.

density, $J_E$, is also shown. In this paper we discuss important aspects of both approaches. We report on the development of a conductive bilayer composed of LaNiO$_3$ and SrRuO$_3$ for Ni-based coated conductors. In addition, we address some stability issues for alloy-based coated conductors by testing transport properties of structures formed by a metal cap layer deposited on YBCO films on single crystal SrTiO$_3$ (STO) substrates.

## II. EXPERIMENTAL

Both SrRuO$_3$ (SRO) and LaNiO$_3$ (LNO) are perovskite-type, conductive metallic oxides with pseudocubic lattice parameters of 3.96 Å and 3.86 Å, and room temperature resistivities of about 300 μΩ-cm and 600 μΩ-cm, respectively. We have previously shown that highly textured LNO buffer layers can be successfully deposited onto Ni tapes [2] and we have subsequently optimized the deposition conditions to reproduce in these films the same degree of in-plane and out-of-plane texture of the Ni substrate [3]. The LNO films for this study were deposited by d.c. sputtering from a powder target of 4" diameter. The previously textured and annealed Ni substrates [4] of approximately 3×15 mm in size were mounted on the heater block, and the chamber evacuated to a base pressure of $1 \times 10^{-5}$ mTorr. After a target presputtering, the temperature was raised to 480 °C and the film grown in a background total pressure $P[(96\%\text{Ar} + 4\%\text{H}_2) + \text{O}_2] = 10$ mTorr, with $P[\text{O}_2]/P[96\%\text{Ar} + 4\%\text{H}_2] = 0.05\text{-}0.1$.

The LNO films show several advantages as a first buffer layer for RABiTS tapes. They can be easily deposited by sputtering at a low temperature compared to that used for other oxides like YSZ and CeO$_2$; LNO deposits on Ni are dense and continuous, with no observed microcracks in films to thicknesses of 0.5 μm. However, YBCO films on LNO/Ni structures show considerably suppressed $T_c$'s (70-75 K) because of Ni diffusion from the substrate though the LNO film into the YBCO [3,5].

Unlike LNO, SRO behaves as a good Ni diffusion barrier and is chemically compatible with YBCO as demonstrated by the growth of YBCO films with $J_c$(0 T, 77 K)=4 MA/cm$^2$ on SRO buffered LaAlO$_3$ single crystals [5]. Since we were not able to grow epitaxial SRO films directly on Ni, SRO was deposited after the first layer of LNO had been sputtered on the Ni substrate. Highly (200) oriented SRO films were deposited by both sputtering and PLD on LNO/Ni samples that had usually been exposed to air after the LNO deposition. For both methods the deposition temperature ranged between 600 and 650 °C. In the case of PLD the deposition was carried out in a background oxygen pressure of 1-5 mTorr and with a laser pulse energy of 4 J/cm$^2$. In the sputtering system SRO was grown using pure Ar at a pressure of 10 mTorr. YBCO films subsequently deposited on these samples showed $J_c$(0 T, 77 K) values around 1.2 MA/cm$^2$ and therefore comparable to those of epitaxial YBCO films on standard, insulating CeO$_2$/YSZ/CeO$_2$ buffered Ni substrate. The YBCO films were deposited by PLD at a substrate temperature of 780 °C and an oxygen background pressure of 100 mTorr.

## III. DISCUSSION

Figure 2 shows a comparison between four-probe resistivity measurements for the Ni substrate, the net SRO/LNO/Ni parallel and the completed YBCO/SRO/LNO/Ni structure.

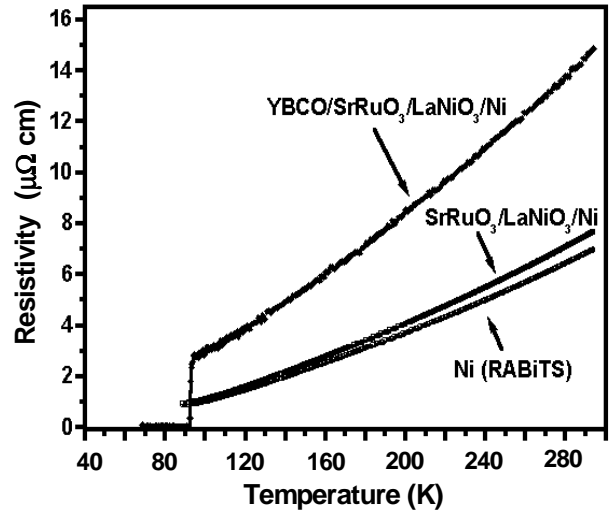

Fig.2. The net resistivity measured using four-terminal potentiometric measurements of: the bare, textured nickel substrate; the nickel tape coated with the conductive bilayer buffer SrRuO$_3$/LaNiO$_3$; and the complete structure of YBCO/SRO/LNO/Ni. The distance between current probes is 10 mm, and the distance between voltage probes is 4 mm.

The thickness values are 140 nm, 150 nm, 325 nm, and 50 μm for YBCO, SRO, LNO and the Ni substrate, respectively. Because of the electrical connection between the YBCO and the Ni substrate through the buffer layers, the resistivity of the YBCO/SRO/LNO/Ni structure is nearly a factor 30 smaller, and the resistance a factor $10^4$ smaller than that of the isolated YBCO film at 100 K. However, were there complete electrical connection between the YBCO film and the Ni substrate the resistivity shown in Fig.2 for the total structure would be indistinguishable from that of the bare Ni substrate. Cross-sectional SEM studies of the YBCO/SRO/LNO/Ni architecture have shown that the additional resistivity is possibly due to a discontinuous NiO layer that forms under the oxidizing conditions of YBCO deposition [3]. Since NiO has a lower free energy of formation than either LNO or SRO, its formation is expected, at a rate that is dependent upon the kinetics of oxygen diffusion to the LNO/Ni interface. The formation of this insulating layer could be explained by the diffusion of Ni from the metal substrate into the LNO layer, thereby resulting in deterioration of the stability of LNO and probably formation of a polycrystalline mixture of La-O and Ni-O. The deterioration of the interface between LNO and Ni and the formation of a NiO layer (although thin and discontinuous) is not desirable for the development of a robust conductive architecture. Ongoing research is focusing on solving this problem by changing the buffer layer architecture or by passivating the Ni surface with a different metal layer.

In the case of coated conductors that use a high resistivity metal substrate, stabilization is carried out by depositing a metal layer on top of the superconducting film. Such a metallic film does not have to grow epitaxially on YBCO, and could be deposited at room temperature. However, as in the case of the conductive buffer layers, the interface between metal and YBCO must have a low areal resistivity in order to provide an effective electrical connection. It is known that the YBCO surface reacts with air to form $Ba(OH)_2$ and $BaCO_3$. Humidity can give rise to other types of surface reconstruction that involve absorption of $H_2O$ molecules inside the YBCO crystal lattice. For this reason, *ex situ* sputter-deposited Ag on the YBCO surface shows surface resistivities that are too high [6,7]. A post-anneal in $O_2$ at 500 °C reduces the contact resistivity to values in the $10^{-5}$ $\Omega$-$cm^2$ range, which may be suitable (note, such a procedure is not possible for Cu on YBCO). However, if the contact surface resistivity of the Ag/YBCO interface is higher than $10^{-8}$-$10^{-9}$ $\Omega$-$cm^2$, then the *I-V* characteristics of our short-segment measurements will show a resistive behavior below $J_c$, with a slope determined by the fraction of current that partitions into the metallic cap layer.

In the best case, the *I-V* characteristic of a metallic layer in parallel with a superconductive layer shows zero voltage up to a current equal to the $I_c$ of the superconductor, a small non-linear transition region for current just above $I_c$, and, finally for $I > I_c$, a linear behavior with a differential resistivity given by that of the metal layer.

To conduct controlled tests of the stabilization provided by a metal cap layer for the case of YBCO insulated from the substrate, we measured *I-V* curves of Ag/YBCO and Cu/Ag/YBCO multilayers grown on insulating single crystal STO. The latter might simulate the ultimate stabilizing structure for coated conductors. After growth, the YBCO films were cooled in a background oxygen pressure of 400 Torr at a rate of 10 °C/min. At room temperature the chamber was evacuated to a background pressure in the range of $10^{-6}$ Torr and a 0.1 μm thick Ag film deposited *in situ* by PLD at a temperature below 100 °C, and using a laser pulse energy of 4 J/$cm^2$ and a repetition rate of 10 to 40 Hz. The samples were then annealed *ex situ* in a furnace at 500 °C for 30 min. Finally, a Ag or Cu film of a few microns in thickness was deposited to complete the stabilizing layer. Cu films were deposited at room temperature by d. c. sputtering using an Ar background of 20 mTorr, with a deposition rate of 30 Å/s.

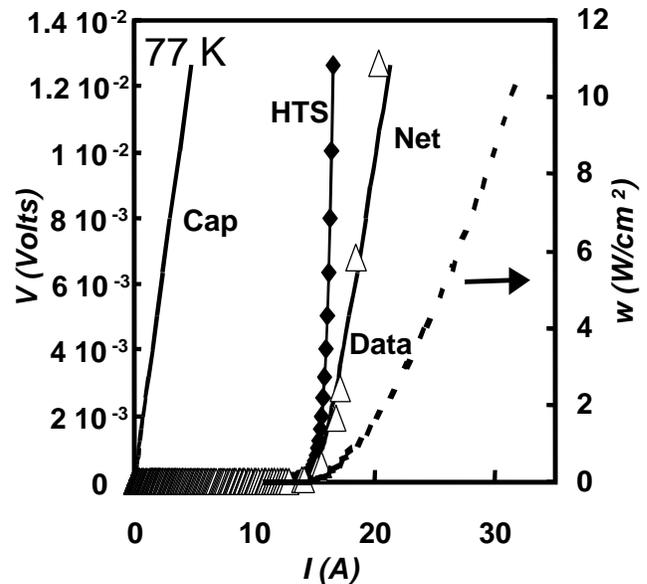

Fig. 3. Comparison between the *I-V* characteristic of a 130 nm isolated YBCO film (diamonds) and the expected *I-V* dependence for the same YBCO film when electrically connected to a 1.5 μm thick Ag film (net solid curve). The dotted line indicates the power per unit area dissipated in the Ag/YBCO bilayer assuming a negligible interface contact resistivity. The open triangles are the experimental data for the Ag/YBCO bilayer. The cap solid line is the Ag film *I-V* characteristic.

Figure 3 is a comparison between the *I-V* curve taken in $LN_2$ for an isolated YBCO film 130 nm thick, and for the same YBCO film after electrically coupled to a 1.5 μm thick Ag cap layer. The substrate dimensions were 3×13 mm, and the YBCO films showed critical current densities in the range of 3 MA/$cm^2$ at 77 K in self-field. In the plot, the solid curves labeled "cap", "HTS", and "net" model the expected partitioning of the total current among the various layers, assuming a negligible interfacial resistance. The data agree well with the expected *I-V* curve for the net, parallel combination of the non-linear superconductor and the ohmic cap layer. The power dissipated in the bilayer is greatly reduced with respect to the isolated YBCO film: 1.7 W/$cm^2$ at $I = 1.7 \times I_c = 20$ A for the bilayer versus ~ 1700 W/$cm^2$ at the same applied current for isolated YBCO. We also notice that the voltage measured in the bilayer below $I_c$ lies in the noise level, indicating a sufficiently low contact resistivity of the Ag/YBCO interface.

Figure 4 shows the same plot generated for a sample formed of a 28 μm Cu film deposited on Ag/YBCO/STO. The Ag thickness is 0.1 μm and the YBCO thickness is 130 nm. In this case the stabilizing effect of the metal layer is much more evident. In fact, the linear differential behavior in the net *I-V* curve is much more pronounced for currents just above $I_c$, and the power dissipated at $2 \times I_c = 21$ A is only 0.24 W/cm$^2$.

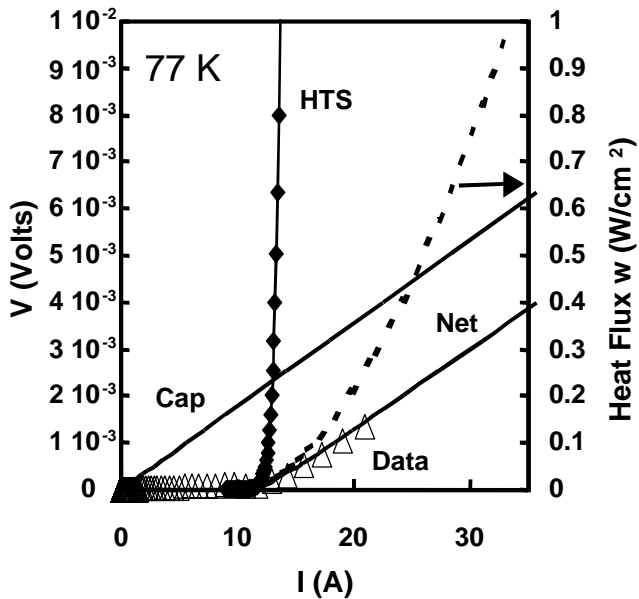

Fig.4. Comparison between the *I-V* characteristic of a 130 nm isolated YBCO film (diamonds) and the expected *I-V* dependence for the same YBCO film when electrically connected to a 28 μm thick Cu film (net solid curve). The dotted line indicates the power per unit area dissipated in the entire structure assuming a negligible interface contact resistivity. The open triangles are the experimental I-V data for the Cu/Ag/YBCO trilayer discussed in the text. The cap solid line is the Cu film *I-V* curve.

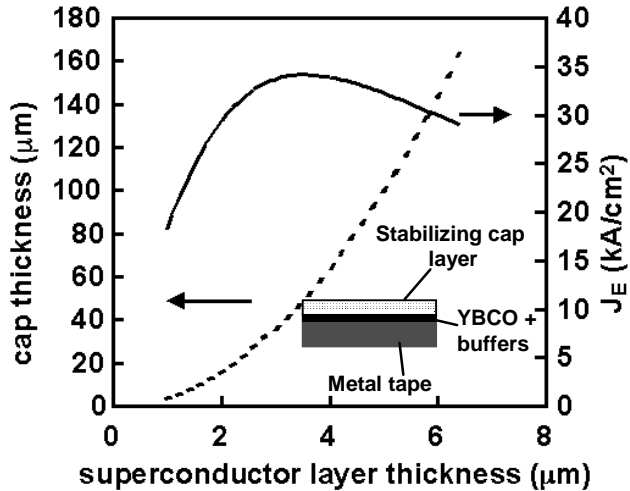

Fig.5. A plot of the calculated cap layer thickness and resulting overall current density, $J_E$, as a function of the HTS layer thickness, for an envisioned coated conductor with a 50 μm thick alloy substrate tape. The curves are generated for the case where $J_c$=1 MA/cm$^2$, and the generated heat flux is limited to 5 W/cm$^2$.

It is worth noting that the requirement of stabilization of a coated conductor (whether it is achieved by a metal cap layer or electrical connection to a low resistivity substrate) and the requirement of maximum overall current density $J_E$ are not independent. The simple model of conduction through parallel layers of different resistivity gives the dependence of $J_E$ on the HTS layer thickness $d_s$ with the condition that, when the superconductor is in the dissipative state ($J_s > J_c$), the power generated in the entire structure is smaller than a criterion value. The calculated $J_E$ shows a maximum $J_E^{max}$ as function of $d_s$. The expression for $J_E^{max}$ and the corresponding superconducting film thickness $d_s^{max}$ are:

$$J_E^{max} = J_c \{(\rho_s - \rho_c)/\rho_s + 2[J_c^2 \rho_c^2 \rho_{m,c}^{-1}(d_m/w)]^{\frac{1}{2}}\}^{-1};$$

$$d_s^{max} = [(w\ d_m)/(J_c^2 \rho_{m,c})]^{\frac{1}{2}},$$

where $\rho_{m,c} = \rho_m \rho_c/(\rho_m - \rho_c)$, $\rho_m$ is the resistivity of the metal substrate, $\rho_c$ is the resistivity of the metal cap layer, $\rho_s$ is the resistivity of the superconductor, $d_m$ is the thickness of the metal substrate, $J_c$ is the critical current density of the superconducting film, and $w$ is the areal power dissipated in the net conductor, respectively. Figure 5 shows a plot of $J_E$ and the thickness of the stabilizing cap layer versus the YBCO thickness, assuming an upper limit of 5 W/cm$^2$ for the dissipated power and a $J_c$ of 1 MA/cm$^2$ for the superconducting layer on an alloy tape. In this case $J_E$ has a maximum value of 34.1 kA/cm$^2$ that corresponds to a YBCO film thickness of 3.5 μm and a metal cap layer thickness of 46.3 μm. Further increase in the YBCO thickness beyond 3.5 μm, while providing more overall current, actually decreases $J_E$ due to the rapid increase in cap layer thickness required to maintain stability.

In conclusion, further development of coated conductor tapes cannot neglect these fundamental considerations. The issues involve the thickness of a metallic stabilizing layer, its interface resistivity with the superconductor, and economic considerations, as required by stability limitations on the power generated in the dissipative regime.